%% file: BPH-10-004_temp.tex
\pdfoutput=1

\documentclass[11pt,twoside,a4paper,cmspaper,final,collab]{cms-tdr}
\def\svnVersion{312277}\def\cmsCernNo{2013-037}\def\cmsCernDate{\today}\def\cmsMessage{Published in Physical Review Letters as \href{http://dx.doi.org/10.1103/PhysRevLett.106.112001}{\doi{10.1103/PhysRevLett.106.112001.}}}

\begin{document}\cmsNoteHeader{BPH-10-004}

\hyphenation{had-ron-i-za-tion}
\hyphenation{cal-or-i-me-ter}
\hyphenation{de-vices}
\RCS$Revision: 40108 $
\RCS$HeadURL: svn+ssh://alverson@svn.cern.ch/reps/tdr2/papers/BPH-10-004/trunk/BPH-10-004.tex $
\RCS$Id: BPH-10-004.tex 40108 2011-02-18 02:28:06Z jolsen $

\def\invnb   {\ensuremath{\mbox{\,nb}^{-1}}}
\def\invpb   {\ensuremath{\mbox{\,pb}^{-1}}}
\def\nb   {\ensuremath{\,\mbox{nb}}}
\def\ub   {\ensuremath{\,\mu\mbox{b}}}
\def\ubNoSpace   {\ensuremath{\mu\mbox{b}}}
\def\um   {\ensuremath{\,\mu\mbox{m}}}
\def\psimumu {\ensuremath{\JPsi\to\Pgmp\Pgmm}}
\def\psik {\ensuremath{\PB^+\to \JPsi\, \PK^+}}
\def\psikCC {\ensuremath{\PB^{\pm}\to \JPsi\, \PK^{\pm}}}
\def\psikst {\ensuremath{B^0\to \JPsi K^{*0}}}
\def\tev  {\ensuremath{\mbox{\,TeV}}}
\def\gev  {\ensuremath{\mbox{\,GeV}}}
\def\gevNospace  {\ensuremath{\mbox{GeV}}}
\def\gevc  {\ensuremath{\mbox{\,GeV}/c}}
\def\gevcc  {\ensuremath{\mbox{\,GeV}/c^2}}
\def\mev  {\ensuremath{\mbox{\,MeV}}}
\def\mevc  {\ensuremath{\mbox{\,MeV}/c}}
\def\mevcc  {\ensuremath{\mbox{\,MeV}/c^2}}
\def\tmu   {\ensuremath{\tt trackerMuon}}
\def\tmus   {\ensuremath{\tt trackerMuons}}
\def\cc   {\ensuremath{c\bar{c}}}
\def\pp {\ensuremath{\Pp\Pp}}
\def\MB {\ensuremath{M_{\PB}}}
\def\ppMuX {{\tt ppMuX}}
\def\dsdpt {\ensuremath{d\sigma/dp_\mathrm{T}^{\PB}}}
\def\dsdeta {\ensuremath{d\sigma/d\eta^{\PB}}}
\def\dsdy {\ensuremath{d\sigma/dy^{\PB}}}
\def\ptb {\ensuremath{p_\mathrm{T}^{\PB}}}
\def\pt {\ensuremath{p_\mathrm{T}}}
\def\etab {\ensuremath{\eta^{\PB}}}
\def\yb {\ensuremath{y^{\PB}}}
\def\ayb {\ensuremath{\left| y^{\PB}\right |}}
\newcommand\T{\rule{0pt}{2.3ex}}
\newcommand\B{\rule[-1.0ex]{0pt}{0pt}}

\cmsNoteHeader{BPH-10-004} 
\title{Measurement of the $\PB^+$ Production Cross Section in \pp\ Collisions at $\sqrt{s} = 7\TeV$}

\date{\today}

\abstract{
Measurements of the total and differential cross sections $\dsdpt$ and $\dsdy$ for
$\PB^+$ mesons produced in \pp\ collisions at $\sqrt{s} = 7\TeV$ are presented.
The data correspond to an integrated luminosity of $5.8\invpb$ collected by the
CMS experiment operating at the LHC.
The exclusive decay \psik, with \psimumu, is used to detect $\PB^+$ mesons
and to measure the production cross section as a function of $\ptb$ and $\yb$.
The total cross section for $\ptb > 5\GeV$ and $\ayb < 2.4$ is
measured to be $28.1\pm 2.4\pm 2.0\pm 3.1\,\mu\mbox{b}$, where the first uncertainty
is statistical, the second is systematic, and the last is from the luminosity
measurement.
}

\hypersetup{
pdfauthor={CMS Collaboration},
pdftitle={Measurement of the B+ Production Cross Section in pp Collisions at sqrts = 7 TeV},
pdfsubject={CMS},
pdfkeywords={CMS, B physics}}

\maketitle 

The study of heavy-quark production in high-energy hadronic interactions
plays a critical role in testing next-to-leading order (NLO)
Quantum Chromodynamics (QCD) calculations~\cite{NDE}.  The first such
measurements were made more than two decades ago by the UA1 Collaboration at
the CERN S$\bar{\Pp}\Pp$S collider~\cite{UA1-87,UA188} operating at a center of
mass energy of $\sqrt{s}=0.63\TeV$, while more recent measurements have
been made by the CDF and D0 Collaborations at the Fermilab Tevatron for $\sqrt{s}=1.8$ and
$1.96\TeV$~\cite{TeVI-CDF1,TeVI-CDF3,TeVI-CDF6,TeVI-D01,TeVI-D03,TeVI-D04,TeVII-CDF1,TEVII-CDF2}.
Substantial progress has been achieved in the understanding of heavy-quark
production at Tevatron energies~\cite{Cacciari04}, but large theoretical
uncertainties remain due to the dependence on the renormalization and
factorization scales.  Particularly important in the perturbative expansion
are terms that scale as powers of $\ln{(\sqrt{s}/m_b)}$ at low transverse
momentum $\pt$ of the $b$ quark~\cite{Collins91,Catani91}, or as powers of
$\ln{(p_\mathrm{T}/m_{b})}$ when
$p_\mathrm{T}\gg m_{b}$~\cite{Cacciari94}, where $m_b$ is the mass of
the $b$ quark.  Measurements of $b$-hadron production at the higher energies provided
by the Large Hadron Collider (LHC) represent an important new test of theoretical
calculations~\cite{Cacciari98,Kniehl08}.

Recently, the LHCb Collaboration measured the production cross section for
$b$ hadrons at the LHC in the forward region using partially reconstructed
decays~\cite{LHCb}.  This Letter presents the first measurement of exclusive 
$\PB$-meson production in \pp\ collisions at $\sqrt{s}=7\TeV$.  A sample of 
$\psikCC$ decays, with $\psimumu$, is reconstructed in $5.84\pm 0.64\invpb$ 
of data collected by the Compact Muon Solenoid (CMS) experiment operating at 
the LHC.  Charge conjugation is assumed in the remainder of this paper, where
$\PB^+$ will be used to refer to both charge states.  The signal yield in bins 
of transverse momentum $\ptb$ and rapidity $\ayb$ is measured with a maximum 
likelihood fit to the reconstructed invariant mass $\MB$ and proper decay length 
$ct$ of the $\PB^+$ candidates.  These yields are corrected for detection 
efficiencies and luminosity to compute the differential production cross sections 
$\dsdpt$ and $\dsdy$.  The results are compared to theoretical predictions based 
on NLO QCD.

A detailed description of the CMS detector can be found elsewhere~\cite{JINST}.
The main subdetectors used in this analysis are the silicon tracker and muon
systems.  The tracker consists of silicon pixel and strip detector modules and is
immersed in a $3.8$\,T magnetic field that
enables the measurement of charged particle momenta over the pseudorapidity range
$|\eta| < 2.5$, where $\eta = -\ln{\tan{\frac{\theta}{2}}}$ and $\theta$ is the
polar angle of the track relative to the counterclockwise beam direction.  
Muons are identified in the range $|\eta|< 2.4$ by gas-ionization detectors
embedded in the steel return yoke.  
The first level of the CMS trigger system consists of custom hardware processors
and uses information from the calorimeters and muon system to 
select the most interesting events in less than $1~\mu$s. The 
High Level Trigger (HLT) processor farm further decreases the event rate to less 
than $300$~Hz before data storage.  
The events used in the measurement reported here were collected with a trigger 
requiring the presence of two muons at HLT with no explicit momentum threshold.

Reconstruction of $\psik$ candidates begins by identifying $\psimumu$
decays.  The muon candidates are required to have at least one reconstructed
segment in the muon system that matches the extrapolated position of a track
reconstructed in the tracker.  Muons within $\left| \eta \right | < 2.4$ that pass
the trigger are selected and further required to satisfy a kinematic threshold that
depends on pseudorapidity:
$\pt^{\Pgm}>3.3\GeV$ for $\left | \eta^{\Pgm}\right | < 1.3$;
$p^{\Pgm}>2.9\GeV$ for $1.3 < \left | \eta^{\Pgm}\right | < 2.2$; and
$\pt^{\Pgm}>0.8\GeV$ for $2.2 < \left | \eta^{\Pgm}\right | < 2.4$.
Candidate $\JPsi$ mesons are reconstructed by combining pairs of oppositely charged
muons having an invariant mass within $150\mev$ of the nominal $\JPsi$
mass~\cite{PDG2010}.  If more than one muon pair in an event satisfies this
selection, the one closest to the $\JPsi$ mass is selected.

Candidate $\PB^+$ mesons are reconstructed by combining a $\JPsi$ candidate with
a track having $\pt>0.9\GeV$, at least four hits in the tracker (of which
one must be in the pixel detector), and a track-fit $\chi^2$ less than five times
the number of degrees of freedom.  A kinematic fit is performed to the dimuon-track
combination, constraining the dimuon mass to equal the $\JPsi$ mass and assuming the third
track to be a kaon.  The selected events must have a resulting $\chi^2$ confidence level
greater than $0.1\%$ and a reconstructed $\PB^+$ mass satisfying $4.95<\MB<5.55\GeV$.
In events with at least one $\PB^+$ candidate, the average number of such candidates
is approximately $1.7$.  When multiple candidates exist, the one with the highest
$\pt$ is retained, which results in the correct choice $95\%$ of the time in simulated
events containing a true signal decay.  A total of $35\,406$ $\PB^+$ candidates pass
all the selection criteria.

The efficiencies corresponding to this selection
range from a few percent for $\ptb\sim 5\GeV$, to approximately $40\%$ for
$\ptb>24\GeV$, as determined in large samples of signal events
generated by \PYTHIA $6.422$~\cite{PYTHIA}, decayed by \EVTGEN~\cite{EvtGen}, and
processed by a detailed simulation of the CMS detector based on \GEANTfour~\cite{GEANT4}.
The efficiencies for hadron-track
reconstruction~\cite{TRK-10-002} and the vertex quality requirement are found to be
consistent between data and simulation within the available precision, which is used
to set the systematic uncertainty of these quantities.  Correction
factors for trigger and muon-reconstruction efficiencies are obtained from a large
sample of inclusive $\psimumu$ decays using a technique similar to that described
in~\cite{BPH-10-002}, where one muon is identified with stringent quality requirements
and the second muon is identified using information separately from the tracker 
or from the muon system.

The proper decay length of each $\PB^+$ candidate is calculated as
$ct = (\MB/\ptb)L_{xy}$, where the transverse decay length $L_{xy}$ is the
vector $\vec{s}$ pointing from the primary vertex~\cite{TRK-10-005} to the
secondary vertex projected onto the $\PB^+$ transverse momentum:
$L_{xy} = (\vec{s}\cdot\vec{p}_T^{\PB})/{\left | \vec{p}_T^B\right |}$.
The core resolution on $ct$ is approximately $30~\mu$m for correctly
reconstructed signal decays.

Backgrounds are dominated by prompt and non-prompt inclusive $\JPsi$ production.
Additional backgrounds arise from misreconstructed b-hadron decays, such as
$\PB\to\JPsi\PKst$, that produce a broad peaking structure in the region
$\MB<5.2\GeV$.   
Contamination from muon pairs that do not originate from $\JPsi$ decay
is negligible after all selection criteria are applied.

The number $n_\text{sig}$ of signal decays in each $\ptb$ and
$\ayb$ bin 
is obtained using an
unbinned extended maximum-likelihood fit to $\MB$ and $ct$.  The likelihood for event
$j$ is obtained by summing the product of yield $n_i$ and probability density ${\cal P}_i$
for each of the signal and background hypotheses $i$.  Five individual components
are considered: signal, $\PB^+\to\JPsi\Pgpp$, misreconstructed $\text{b}\bar{\text{b}}$ 
events that peak in $\MB$, non-prompt $\JPsi$, and prompt $\JPsi$.  The extended
likelihood function is then the product of likelihoods for all events:
\begin{equation}
{\cal L} = \exp\left ( -\sum_i n_i \right ) \prod_j \left [
\sum_i n_i{\cal P}_i(\MB;\, \vec{\alpha}_i){\cal P}_i(ct;\, \vec{\beta}_i) \right ].
\end{equation}
The probabilities ${\cal P}_i$ are the probability density
functions (PDFs) with shape parameters $\vec{\alpha}_i$ for $\MB$, and
$\vec{\beta}_i$ for $ct$, evaluated separately for each of the $i$ fit components.
The yields $n_i$ are then determined by maximizing ${\cal L}$ with
respect to the yields and a subset of the PDF parameters.  The yield for
$\JPsi\Pgpp$ is constrained to equal the $\JPsi\,\PK^+$ yield times the ratio
of branching fractions for the two decay modes~\cite{PDG2010}.  

The $\MB$ PDFs are the sum of three (two)
Gaussians for the signal ($\JPsi\Pgp$) with parameters obtained
from simulation; an exponential for both prompt and non-prompt
$\JPsi$;
and a combination of two Gaussians and an exponential for the peaking
$b\bar{b}$ background.  The resolution on $\MB$ for signal decays is approximately 
$30\MeV$.  The $ct$ PDFs are a single exponential convolved with the resolution
function to describe the signal, $\JPsi\Pgp$, and peaking background components, where
the lifetime is allowed to be different for the latter; the sum of two exponentials
convolved with the resolution function for the non-prompt $\JPsi$ component; and the
pure resolution function for the prompt $\JPsi$ component. The resolution function is
common for signal and background, and is described by the sum of two or three Gaussian
functions, depending on $\ptb$ and $\ayb$.

\begin{table*}[tb]
\caption{Bin ranges for $\ptb$ and $\ayb$, signal yields $n_{\rm sig}$,
efficiencies $\epsilon$, and measured differential cross sections $\dsdpt$
and $\dsdy$, compared to the \MCATNLO~\cite{MCNLO} and \PYTHIA predictions.
The uncertainties in the measured cross sections are statistical and systematic, respectively,
excluding the common branching fraction ($3.5\%$) and luminosity ($11\%$) uncertainties.
The result for $\ptb>30\GeV$ is quoted as an integrated cross section in $\ubNoSpace$.}
\label{tab:results}
\begin{center}
\begin{tabular*}{1.0\textwidth}{@{\extracolsep{\fill}}cccccc} \hline
$\ptb\,(\text{GeV})$ \T \B      & $n_\text{sig}$ & $\epsilon\,(\%)$     & $\dsdpt\,(\ubNoSpace/\text{GeV})$	& \MCATNLO & \PYTHIA \\ \hline
$5$--$10$ \T \B         & $223 \pm 26$  & $1.56\pm 0.02$ & $4.07  \pm 0.47 \pm 0.31$     & $3.72^{+1.46}_{-0.89}$    & $6.68$ \\
$10$--$13$ \T \B        & $236 \pm 21$  & $7.62\pm 0.11$ & $1.47  \pm 0.13 \pm 0.09$     & $1.17^{+0.31}_{-0.24}$    & $2.66$ \\
$13$--$17$  \T \B       & $169 \pm 17$  & $14.6\pm 0.2$  & $0.412 \pm 0.041 \pm 0.026$   & $0.47^{+0.10}_{-0.05}$    & $1.01$ \\
$17$--$24$  \T \B       & $207 \pm 17$  & $23.3\pm 0.6$  & $0.181 \pm 0.015 \pm 0.012$   & $0.15^{+0.04}_{-0.03}$    & $0.28$ \\
$24$--$30$ \T \B        & $56  \pm 9$   & $31.9\pm 1.5$  & $0.042 \pm 0.007 \pm 0.004$   & $0.048^{+0.029}_{-0.018}$ & $0.08$ \\
$>30$ \T \B	        & $44  \pm 8$   & $33.4\pm 2.0$  & $0.188 \pm 0.034 \pm 0.018$ & $0.20^{+0.11}_{-0.02}$    & $0.27$ \\\hline
$\ayb$  \T \B     & $n_\text{sig}$ & $\epsilon\,(\%)$     & $\dsdy\,(\mu\text{b})$	 & \MCATNLO & \PYTHIA \\\hline
$0.00$--$0.60$ \T \B    & $187 \pm 17$  & $3.01\pm 0.06$ & $7.39 \pm 0.65 \pm 0.53$	 & $5.98^{+2.2}_{-1.31}$    & $11.1$ \\
$0.60$--$1.10$ \T \B    & $164 \pm 17$  & $3.81\pm 0.08$ & $6.11 \pm 0.64 \pm 0.47$	 & $5.85^{+1.78}_{-1.37}$    & $10.8$ \\
$1.10$--$1.45$\T \B     & $207 \pm 20$  & $5.92\pm 0.12$ & $7.11 \pm 0.69 \pm 0.59$	 & $5.59^{+1.71}_{-1.31}$    & $10.2$ \\
$1.45$--$1.80$ \T \B    & $203 \pm 22$  & $8.24\pm 0.15$ & $5.01 \pm 0.55 \pm 0.42$	 & $4.96^{+1.88}_{-1.10}$    & $9.5$  \\
$1.80$--$2.40$ \T \B & $176 \pm 22$  & $6.31\pm 0.12$ & $3.31 \pm 0.42 \pm 0.28$	 & $4.29^{+1.73}_{-1.14}$    & $8.5$  \\\hline
\end{tabular*}
\end{center}
\end{table*}

The fit proceeds in several steps so that all background shapes are obtained
directly from data, except for the peaking component.  This technique relies on the
assumption that in the signal-free region $5.40 < \MB < 5.55\GeV$ (upper sideband) there
are only two contributions: prompt and non-prompt $\JPsi$ background (ignoring the
small contribution from $\JPsi\Pgp$).  To obtain the effective lifetime
of the non-prompt $\JPsi$ background, the $ct$ distribution is fitted for events in the
inclusive $\PB^+$ sample defined by $\ptb>5\GeV$ and $\ayb < 2.4$ that lie in the $\MB$ 
upper sideband region, allowing the resolution function parameters to vary freely.  The 
resolution function is then fixed and the signal $\PB^+$ lifetime in the inclusive sample 
is obtained by fitting $ct$ and $\MB$ simultaneously.  The result, 
$c\tau = 481\pm 22\,\mu$m (statistical uncertainty only), is in good agreement with the
world-average value of $491\pm 9\,\mu$m~\cite{PDG2010}.  With the effective lifetime for
signal and non-prompt background fixed, the resolution function parameters are then determined 
separately in each bin of $\ptb$ and $\ayb$.  Finally, with all $ct$ resolution and background 
lifetime parameters fixed, the signal and background yields are fitted in each bin, together 
with the parameters describing the shape of the prompt and non-prompt $\JPsi$ components in $\MB$.

The accuracy and robustness of the fit strategy were checked with a set of
$400$ pseudoexperiments where signal and background
events were generated randomly from the PDFs in each bin.  The fitted yields were
unbiased and the uncertainties were estimated properly.  
The effects of correlations between $\MB$ and $ct$ were studied by mixing together fully 
simulated signal and background events to produce $100$ pseudoexperiments.  No significant evidence 
of bias in the signal yield was found, and the observed deviations
(a few percent) between fitted and generated yields are taken as the systematic uncertainty
due to potential biases in the fit method.

\begin{figure}[tb]
  \begin{center}
   \includegraphics[height=0.95\columnwidth,angle=90]{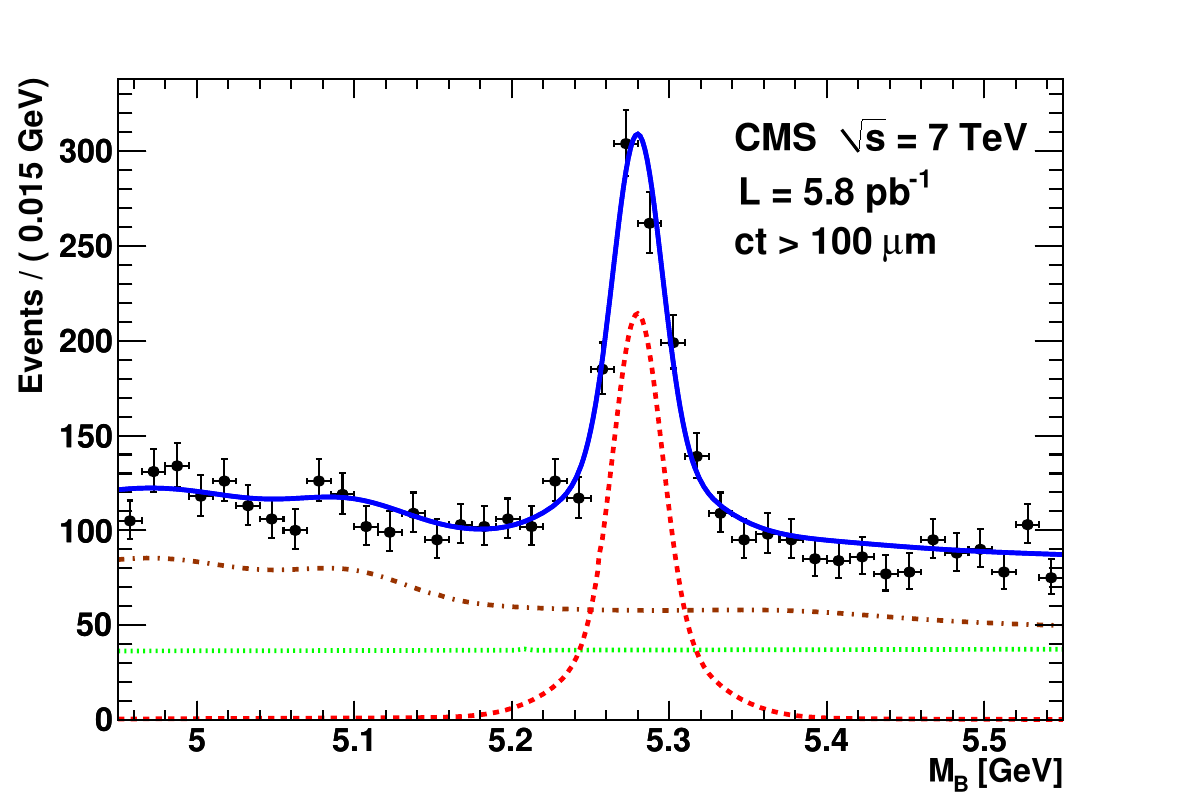}
   \includegraphics[width=0.95\columnwidth]{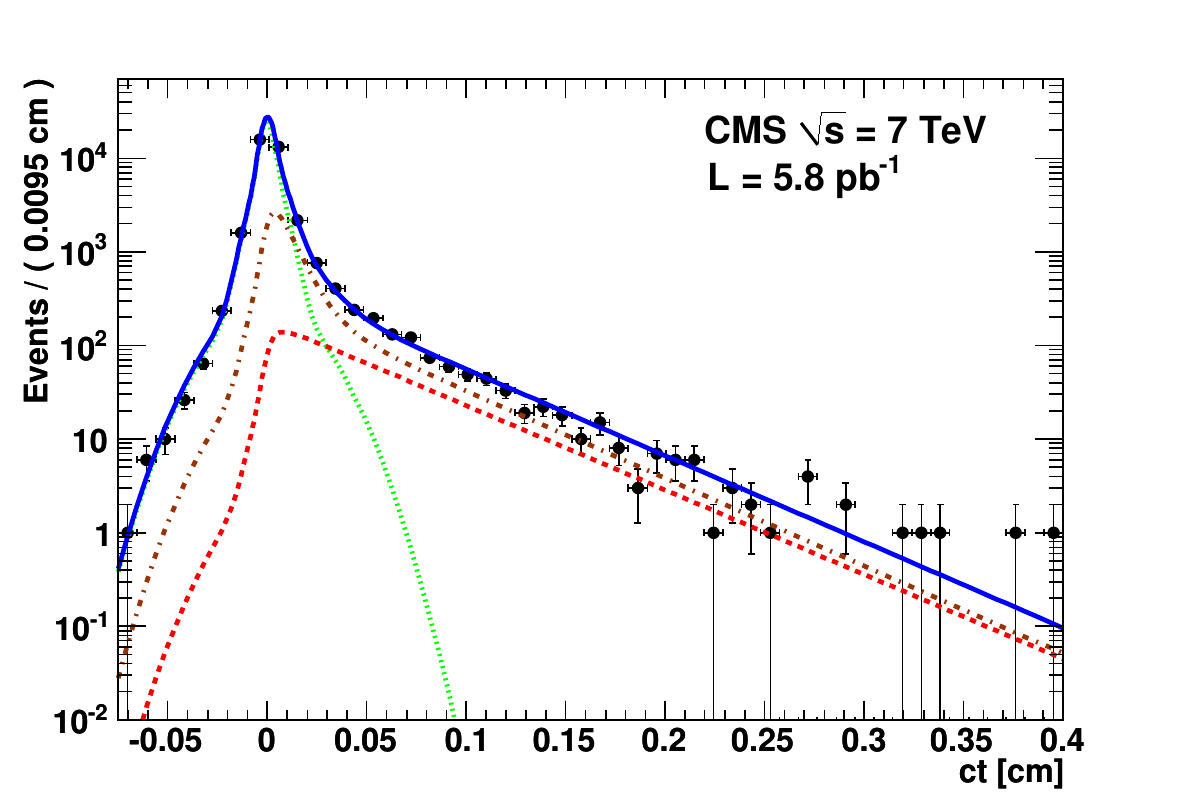}
    \caption{Projections of the fit results in $\MB$ (top) and $ct$ (bottom) for
    $\ptb>5\GeV$ and $\ayb < 2.4$.  The curves in
    each plot are the sum of all contributions (solid blue line); signal (dashed red);
    prompt $\JPsi$ (dotted green); and the sum of non-prompt $\JPsi$, peaking $b\bar{b}$,
    and $\JPsi\Pgpp$ (dot-dashed brown).  For better visibility of the individual contributions,
    the $\MB$ plot includes a requirement of $ct>100\,\mu{\rm m}$.}
    \label{fig:fit}
  \end{center}
\end{figure}

Table~\ref{tab:results} summarizes the fitted signal yield in each bin of $\ptb$ and
$\ayb$, while Fig.~\ref{fig:fit} shows the fit projections for $\MB$ and $ct$ from the
inclusive sample with $\ptb>5\GeV$ and $\ayb < 2.4$.  The total number of signal events
is $912\pm 47$, where the error is statistical only.

The differential cross sections for $\PB^+$ production as a function of $\ptb$ and
$\yb$ (averaged for positive and negative rapidities) are defined as
\begin{equation}
\begin{aligned}
\frac{d\sigma(\Pp\Pp\to \PB^+ X)}{dp_\mathrm{T}^{\PB}} &= \frac{n_{\rm sig}(\ptb)}
{2\,\epsilon(\ptb) \,{\cal B}\,\lumi\,\Delta \ptb},\\
\frac{d\sigma(\Pp\Pp\to \PB^+ X)}{d\yb} &= \frac{n_{\rm sig}(\ayb)}
{2\,\epsilon(\ayb) \,{\cal B}\,\lumi\,\Delta \yb},
\label{eq:Xsec}
\end{aligned}
\end{equation}
where $n_\text{sig}(\ptb)$ and $n_\text{sig}(\ayb)$ are the fitted signal
yields in the given bin, $\epsilon(\ptb)$ and $\epsilon(\ayb)$ are the efficiencies
in each bin for a $\PB^+$ meson produced with $\ptb>5\GeV$ and $\ayb<2.4$ to pass all
the selection criteria, $\Delta \ptb$ is the bin size in $\ptb$, and
$\Delta \yb = 2\,\Delta\!\ayb$ is the bin size in $\yb$.  The total branching fraction
${\cal B}$ is the product of
the individual branching fractions ${\cal B}(\psik) = (1.014\pm 0.034)\times 10^{-3}$ and
${\cal B}(\psimumu)=(5.93\pm 0.06)\times 10^{-2}$~\cite{PDG2010}.
The factor of two in the denominator of Eq.~\ref{eq:Xsec} takes into account the
choice of quoting the cross section for a single charge (taken to be $\PB^+$), while
$n_{\rm sig}$ includes both charge states.  All efficiencies, $\epsilon(\ptb)$ or
$\epsilon(\ayb)$, are calculated separately in each bin,
and account for bin-to-bin migrations (a few percent) due to the
resolution on the measured momentum and rapidity.

\begin{figure}[!tb]
\begin{center}
\includegraphics[width=0.70\columnwidth,angle=90]{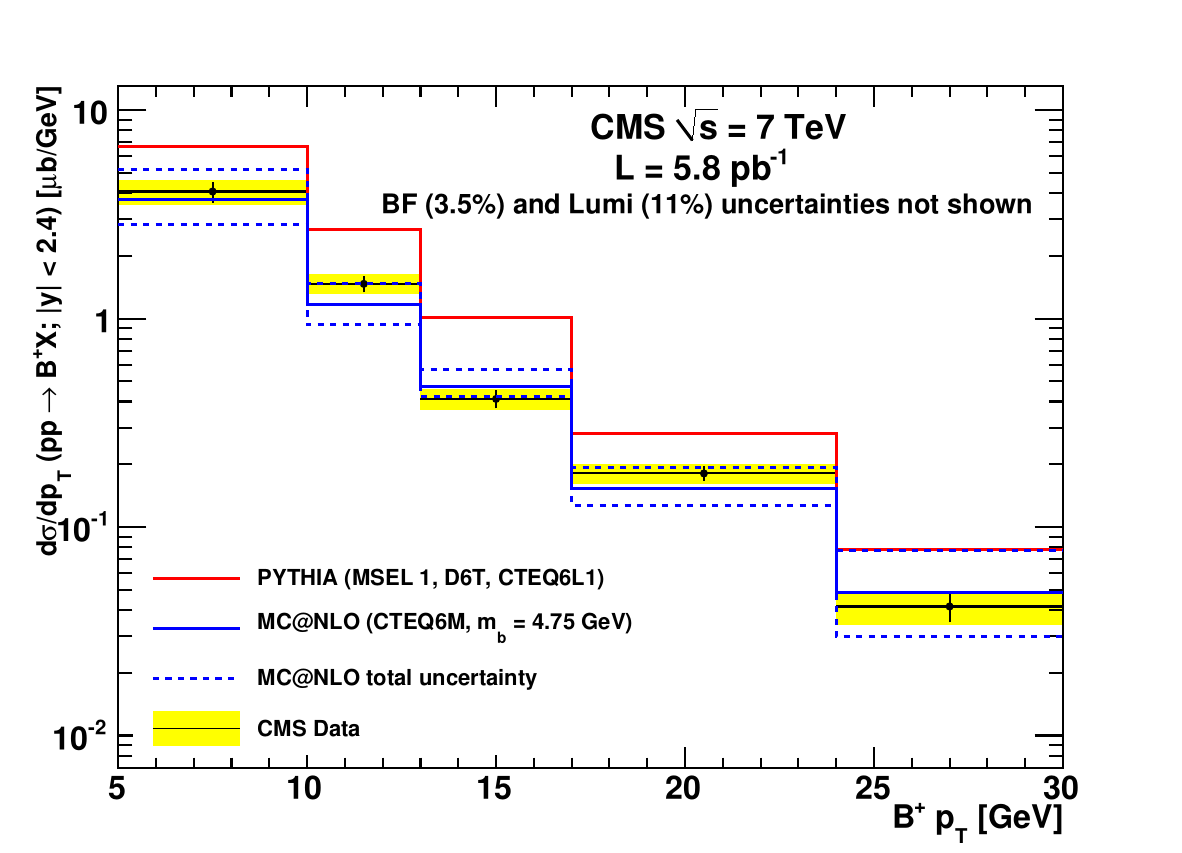}
\includegraphics[width=0.70\columnwidth,angle=90]{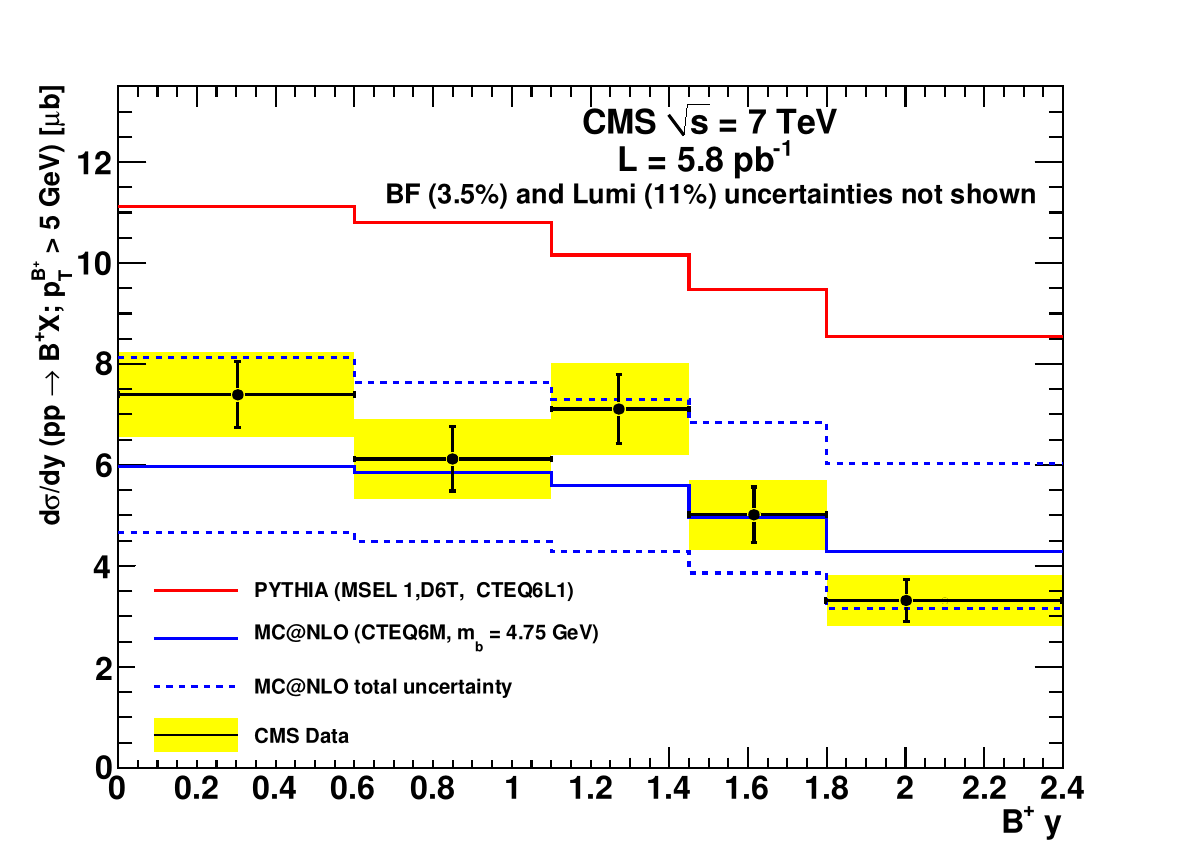}
\caption{Measured differential cross sections
$\dsdpt$ (top) and $\dsdy$ (bottom) compared with the
theory predictions.  The error bars are the statistical uncertainties,
while the (yellow) band represents the sum in quadrature of statistical and
systematic uncertainties, excluding the common branching fraction
and luminosity uncertainties.
The solid and dashed blue lines are the \MCATNLO prediction and its uncertainty,
respectively.  The solid red line is the \PYTHIA prediction.}
\label{fig:Xsec}
\end{center}
\end{figure}

The cross section is affected by several sources of systematic uncertainty arising
from the signal yields, efficiencies, branching fractions, and luminosity.  Uncertainties
of the signal yields arise from potential fit biases and imperfect knowledge of the PDF
parameters ($2$--$5\%$), $ct$ resolution function ($1$--$2\%$), and the effects of final-state
radiation on the signal shape in $\MB$ ($<1\%$).  Uncertainties of the trigger ($2\%$), muon
identification ($1\%$), and tracking ($1$--$4\%$) efficiencies are all determined directly
from data.  The contribution ($1$--$4\%$) related to the $\PB^+$ momentum spectrum is
evaluated by reweighting the shape of the $\ptb$
distribution generated with \PYTHIA to match
the spectrum predicted by \MCATNLO $3.4$~\cite{MCNLO}.
An uncertainty of $1.5\%$ is assigned to the efficiency of the vertex quality requirement.
The effect of tracker misalignment on the cross sections due to
variations in the signal yields and efficiencies is estimated to be approximately
$2\%$ using samples simulated with a different alignment than the nominal one.  The total
systematic uncertainty of the cross section measurement in each bin is computed as the sum
in quadrature of the individual uncertainties, and is summarized in Table~\ref{tab:results}.
In addition, there are common uncertainties of $3.5\%$ from the branching fractions and $11\%$
from the luminosity measurement~\cite{EWK-10-004}.

The differential cross sections as functions of $\ptb$ and $\ayb$ are shown in
Fig.~\ref{fig:Xsec} and Table~\ref{tab:results}.  They are compared with the predictions of
\MCATNLO using a $b$-quark mass of $4.75\GeV$, renormalization and factorization scales
$\mu=\sqrt{m_b^2+p_\mathrm{T}^2}$, and the CTEQ6M parton distribution
functions~\cite{CTEQ}.  The uncertainty on the predicted cross section is calculated by
varying the renormalization and factorization scales by a factor of two, $m_b$
by $\pm 0.25\GeV$, and by using the CTEQ$6.6$ parton distribution set.
For reference, the prediction of \PYTHIA is also included, using a $b$-quark mass of
$4.8\GeV$, CTEQ6L1 parton distributions~\cite{CTEQ}, and the D6T tune~\cite{D6T} to simulate the
underlying event.  The total integrated cross section for $\ptb>5\GeV$ and $\ayb < 2.4$ is
calculated as the sum over all $\ptb$ bins and is found to be
$28.1\pm 2.4\pm 2.0\pm 3.1\ub$, where the first uncertainty is statistical, the second is 
systematic (including the branching fraction uncertainty), and the last is from the luminosity 
measurement.  
This result lies between the predictions of \MCATNLO,
$25.5^{+8.8}_{-5.4}\,\text{(scale)}^{+2.5}_{-1.8}\,\text{(mass)}\pm 0.8\,\text{(PDF)}\ub$,
and \PYTHIA ($48.1~\mu$b).

In summary, first measurements of the total and differential cross sections for
charged $\PB$ production in $\Pp\Pp$ collisions at $\sqrt{s}=7\TeV$ using the decay $\psikCC$ 
have been presented.  The measurements cover the range $\ayb < 2.4$ and $\ptb$ from $5\GeV$ to 
greater than $30\GeV$.  The result is in reasonable agreement with the predictions of 
\MCATNLO\ in terms of shape and absolute normalization.

We wish to congratulate our colleagues in the CERN accelerator departments for the
excellent performance of the LHC machine. We thank the technical and administrative
staff at CERN and other CMS institutes, and acknowledge support from: FMSR (Austria);
FNRS and FWO (Belgium); CNPq, CAPES, FAPERJ, and FAPESP (Brazil); MES (Bulgaria); CERN;
CAS, MoST, and NSFC (China); COLCIENCIAS (Colombia); MSES (Croatia); RPF (Cyprus);
Academy of Sciences and NICPB (Estonia); Academy of Finland, ME, and HIP (Finland);
CEA and CNRS/IN2P3 (France); BMBF, DFG, and HGF (Germany); GSRT (Greece);
OTKA and NKTH (Hungary); DAE and DST (India); IPM (Iran); SFI (Ireland); INFN (Italy);
NRF and WCU (Korea); LAS (Lithuania); CINVESTAV, CONACYT, SEP, and UASLP-FAI (Mexico);
PAEC (Pakistan); SCSR (Poland); FCT (Portugal); JINR (Armenia, Belarus, Georgia,
Ukraine, Uzbekistan); MST and MAE (Russia); MSTD (Serbia); MICINN and CPAN (Spain);
Swiss Funding Agencies (Switzerland); NSC (Taipei); TUBITAK and TAEK (Turkey);
STFC (United Kingdom); DOE and NSF (USA).

\bibliography{auto_generated}	
\cleardoublepage \appendix\section{The CMS Collaboration \label{app:collab}}\begin{sloppypar}\hyphenpenalty=5000\widowpenalty=500\clubpenalty=5000\input{BPH-10-004-authorlist.tex}\end{sloppypar}
\end{document}

%% file: BPH-10-004-authorlist.tex
\textbf{Yerevan Physics Institute,  Yerevan,  Armenia}\\*[0pt]
V.~Khachatryan, A.M.~Sirunyan, A.~Tumasyan
\vskip\cmsinstskip
\textbf{Institut f\"{u}r Hochenergiephysik der OeAW,  Wien,  Austria}\\*[0pt]
W.~Adam, T.~Bergauer, M.~Dragicevic, J.~Er\"{o}, C.~Fabjan, M.~Friedl, R.~Fr\"{u}hwirth, V.M.~Ghete, J.~Hammer\cmsAuthorMark{1}, S.~H\"{a}nsel, C.~Hartl, M.~Hoch, N.~H\"{o}rmann, J.~Hrubec, M.~Jeitler, G.~Kasieczka, W.~Kiesenhofer, M.~Krammer, D.~Liko, I.~Mikulec, M.~Pernicka, H.~Rohringer, R.~Sch\"{o}fbeck, J.~Strauss, A.~Taurok, F.~Teischinger, P.~Wagner, W.~Waltenberger, G.~Walzel, E.~Widl, C.-E.~Wulz
% ...
\vskip\cmsinstskip
% ...
1:~~Also at CERN, European Organization for Nuclear Research, Geneva, Switzerland\\
% ...
